\documentclass[preprint,showpacs,preprintnumbers,amsmath,amssymb,floatfix,endfloats*]{revtex4}

\usepackage{graphicx}
\usepackage{dcolumn}
\usepackage{bm}
\usepackage{amsfonts}

\DeclareMathAlphabet{\mathsfsl}{OT1}{cmr}{bx}{it}
\begin{document}
\title{Diffusion of a Janus nanoparticle in an explicit solvent:
A molecular dynamics simulation study}
\author{Ali Kharazmi$^{1}$ and Nikolai V. Priezjev$^{2}$}
\affiliation{$^{1}$Department of Mechanical Engineering, Michigan
State University, East Lansing, Michigan 48824}
\affiliation{$^{2}$Department of Mechanical and Materials
Engineering, Wright State University, Dayton, Ohio 45435}
\date{\today}
%
\begin{abstract}

Molecular dynamics simulations are carried out to study the
translational and rotational diffusion of a single Janus particle
immersed in a dense Lennard-Jones fluid.    We consider a spherical
particle with two hemispheres of different wettability.   The
analysis of the particle dynamics is based on the time-dependent
orientation tensor, particle displacement, as well as the
translational and angular velocity autocorrelation functions.  It
was found that both translational and rotational diffusion
coefficients increase with decreasing surface energy at the
nonwetting hemisphere, provided that the wettability of the other
hemisphere remains unchanged.   We also observed that in contrast to
homogeneous particles, the nonwetting hemisphere of the Janus
particle tends to rotate in the direction of the displacement vector
during the rotational relaxation time.

\end{abstract}

\pacs{68.08.-p, 66.20.-d, 83.10.Rs}


\maketitle

\section{Introduction}

Understanding the processes of self- and directed-assembly of
nanoparticles is important for design of various nanostructured
materials with advantageous mechanical and optical
properties~\cite{Israel08}.   Recent progress in synthesis of Janus
and asymmetric patchy particles, that can form a variety of
predictable
superstructures~\cite{Glotzer04,Luijten08,Gubbins12,An12}, makes
their production feasible at an industrial
scale~\cite{Granick10,Loget12}.    The dynamics of assembly depends
on the particle structure, distribution of wettability and charges,
and boundary conditions at the particle surface in a solvent. The
hydrodynamic boundary conditions are usually specified via the
so-called slip length, which is defined as an extrapolated distance
with respect to the interface where the velocity profile
vanishes~\cite{deGennes85}. In general, it was shown that the slip
length is a tensor quantity for flows over anisotropic textured
surfaces~\cite{Vinograd08,Priezjev05,Vinograd11,Priezjev11}.
Transport properties of chemically homogeneous nanoparticles were
recently studied in cylindrical pores filled with a wetting
fluid~\cite{Drazer05} and in an atomistic solvent confined between
flat walls~\cite{Khare10}.     However, the diffusion process of
individual Janus particles even in the absence of flow or
confinement remains not fully understood.

\vskip 0.05in

At the continuum level, the hydrodynamic behavior of Janus particles
was studied in the low Reynolds number
regime~\cite{Swan08,Willmott08,Willmott09,Chan13}. It was found that
when the slip length is smaller than the particle size, the
translational velocity of the spherical Janus particle is coupled to
the moments of the force density on its surface~\cite{Swan08}.
Moreover, it was shown that due to its asymmetrical boundary
conditions, Janus particles can migrate parallel to the velocity
gradient in shear flows~\cite{Swan08}.     In the limit of small
slip lengths, the torque and force on spherical Janus particles with
either discontinuous binary or continuously patterned surfaces were
computed~\cite{Willmott08, Willmott09}.   More recently, torques and
forces on pill-shaped Janus particles with stick and slip boundary
conditions were analyzed using the boundary integral
method~\cite{Chan13}.   In particular, it was demonstrated that
depending on the aspect ratio and slip length, the force transverse
to the direction of a uniform flow can change sign~\cite{Chan13}.
In the present work, the diffusive motion of a Janus particle in a
quiescent fluid was studied using molecular dynamics (MD)
simulations.

\vskip 0.05in

In recent years, Brownian motion of spherical particles with {\it
uniform} wettability was extensively studied by MD
simulations~\cite{Levesque00,Schmidt03,Kapral04,Levesque07,Li09,Shin10,Chakraborty11,Rudyak11}.
It was shown that a number of parameters, e.g., the mass and size
ratios of a particle and a solvent molecule, interaction potentials,
boundary conditions and the system size, can influence the diffusion
process.    Notably, at long times, the characteristic algebraic
decay of the velocity autocorrelation function (VACF) due to
hydrodynamic correlation between a particle and a
fluid~\cite{Alder70} depends on the particle
diameter~\cite{Levesque07}; while at short times, the VACF exhibits
pronounced oscillations at small values ($\lesssim 10$) of the mass
ratio between the particle and the fluid
molecule~\cite{Levesque00,Shin10}.     It was also found that when
the ratio of the system size to the particle radius is larger than
about $20$, then the finite size effects due to particle images are
not important~\cite{Li09,Chakraborty11}.    Furthermore, in the case
of a smooth spherical particle, the Stokes-Einstein relation for the
diffusion coefficient holds for the system with slip boundary
conditions~\cite{Levesque00}.   However, if the particle consists of
a cluster of atoms and their interaction energy with fluid molecules
is sufficiently large, then stick boundary conditions apply at the
particle surface~\cite{Li09,Chakraborty11}.

\vskip 0.05in

In this paper, the results of molecular dynamics simulations are
reported for the translational and rotational diffusion of
individual Janus nanoparticles in a quiescent solvent.   In our
study, Janus particles are modeled as spheres with two sides of
different wettability, and their behavior is contrasted with that of
uniformly wetting and nonwetting particles.    We find that both
translational and rotational diffusion coefficients are larger for
Janus particles with lower surface energy at the nonwetting
hemisphere, provided that the wettability of the other hemisphere
remains the same.     It also will be shown that long-time algebraic
decay of the velocity autocorrelation functions is recovered for
uniform and anisotropic particles.     Interestingly, the diffusive
motion of Janus particles involves a subtle correlation between
translational and angular velocities; i.e., a non-zero velocity
component parallel to the particle symmetry plane causes a finite
rotation to reduce drag.

\vskip 0.05in

The rest of the paper is structured as follows.  The description of
the molecular dynamics simulation model and parameter values is
given in the next section.   The analysis of the fluid density
profiles, boundary conditions, particle displacement and rotation,
as well as translational and angular velocity autocorrelation
functions is presented in Sec.\,\ref{sec:Results}.   The conclusions
are provided in the last section.

\section{Molecular dynamics simulation model}
\label{sec:MD_Model}

The diffusive motion of a Janus particle in an explicit solvent was
studied using classical molecular dynamics
simulations~\cite{Lammps}.   In the explicit solvent model, the
pairwise interaction between fluid monomers is specified via the
truncated Lennard-Jones (LJ) potential
\begin{equation}
V_{LJ}(r)=4\,\varepsilon\,\Big[\Big(\frac{\sigma}{r}\Big)^{12}\!-\Big(\frac{\sigma}{r}\Big)^{6}\,\Big],
\label{Eq:LJ}
\end{equation}
where $\varepsilon$ is the depth of the potential well and $\sigma$
is the separation distance at which the LJ potential is zero.    The
cutoff radius was fixed to $r_{c}=2.5\,\sigma$ for all types of
interactions.    The system consists of $46\,531$ fluid monomers and
a Janus particle, which are confined into a cubic box of linear size
$39.62\,\sigma$.      All simulations were performed at a constant
volume with the fluid density $\rho=0.75\,\sigma^{-3}$ and
temperature $1.1\,\varepsilon/k_B$, where $k_B$ is the Boltzmann
constant.   The fluid temperature was maintained by the
Nos\'{e}-Hoover thermostat with the coupling parameter of $1.0$.
Periodic boundary conditions were applied in all three directions.
The equations of motion were integrated numerically using the Verlet
algorithm~\cite{Allen87} with a time step $\triangle
t_{MD}=0.005\,\tau$, where $\tau=\sigma\sqrt{m/\varepsilon}$ is the
time scale of the LJ potential.

\vskip 0.05in


The Janus particle was modeled as a convex polyhedron, the so-called
Pentakis snub dodecahedron, which consists of $140$ triangular
faces, $210$ edges, and $72$ vertices~\cite{Conway08}.  Out of the
total $210$ edges, there are $60$ short ($0.86\,\sigma$) and $150$
long ($0.99\,\sigma$) edges.     The radius of a sphere that
contains all vertices is $2.08\,\sigma$.    In our model, 72 LJ
atoms were fixed relative to each other at the vertices of the
polyhedron; and, during the simulations, the Janus particle was
moving as a rigid body under forces from the surrounding fluid. The
particle atoms do not interact with each other and their size is the
same as the size of fluid monomers.     The particle mass is
$M=50.0\,m$.   The surface of the Janus particle was divided into
two hemispheres of different wettability, as shown in
Fig.\,\ref{fig:snapshot_janus}.    The LJ interaction energy between
fluid monomers and particle atoms at the wetting hemisphere was
fixed to $\varepsilon_{\rm pf} = 1.0\,\varepsilon$, and at the
nonwetting hemisphere $\varepsilon_{\rm pf}$ was varied in the range
$0.1\,\varepsilon \leqslant \varepsilon_{\rm pf} \leqslant
1.0\,\varepsilon$.  The two limiting cases of \textit{uniformly}
wetting and nonwetting particles correspond to the values of the
interaction energy with the fluid molecules of $1.0\,\varepsilon$
and $0.1\,\varepsilon$, respectively.

\vskip 0.05in


The Janus particle and the surrounding fluid phase were first
equilibrated for about $2\times10^7$ MD steps (or $10^5\,\tau$) and
then the data were accumulated during the next $6\times10^8$ MD
steps (or $3\times 10^6\,\tau$).     During the production runs, the
position and orientation of the Janus particle, the velocity of all
particle atoms as well as the position of the center of mass of the
solvent were saved every $20$ MD steps.     To accumulate
statistics, the simulations were repeated in forty independent
systems for each value of $\varepsilon_{\rm pf}$.   In each case,
the velocities of fluid monomers were initialized using the
Maxwell-Boltzmann probability distribution with zero total linear
and angular momenta~\cite{Lammps}.   We also checked that after
equilibration, the Maxwell-Boltzmann distribution for the
translational and angular velocities of the Janus particle is
recovered.

\section{Results}
\label{sec:Results}

\begin{figure}[t]
\includegraphics[width=12.cm,angle=0]{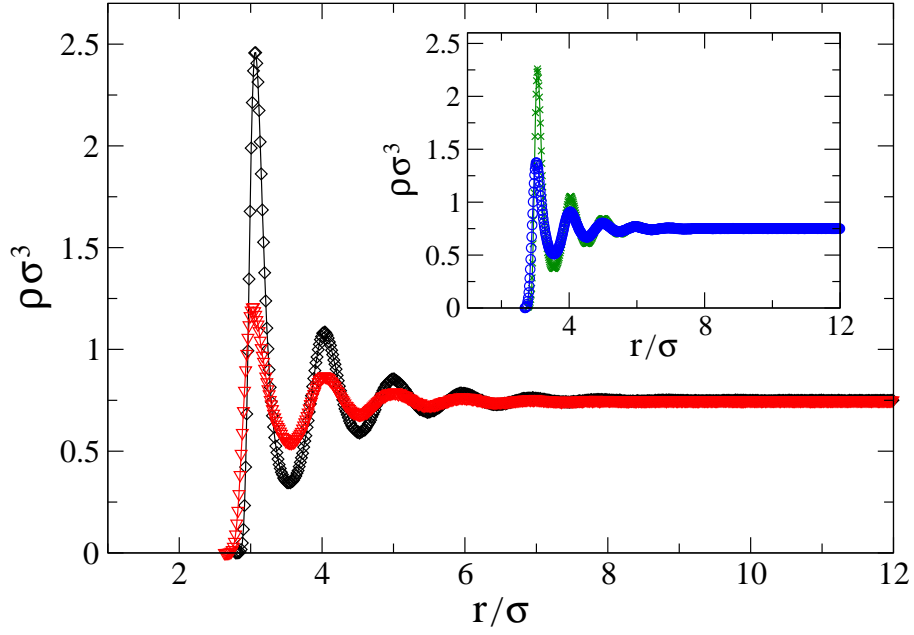}
\caption{(Color online) Averaged fluid density profiles as a
function of the distance from the particle center for
\textit{uniformly} wetting ($\diamond$) and \textit{uniformly}
nonwetting ($\triangledown$) particles.  The LJ interaction energy
between particle atoms and fluid monomers is $\varepsilon_{\rm
pf}=1.0\,\varepsilon$ ($\diamond$) and $\varepsilon_{\rm
pf}=0.1\,\varepsilon$ ($\triangledown$).   The inset shows fluid
density profiles near the wetting hemisphere $\varepsilon_{\rm
pf}=1.0\,\varepsilon$ ($\times$) and nonwetting hemisphere
$\varepsilon_{\rm pf}=0.1\,\varepsilon$ ($\circ$) of the Janus
particle. } \label{fig:density_radial}
\end{figure}

It has long been recognized that in the presence of a smooth
substrate, the fluid monomers tend to form several distinct layers
parallel to the substrate~\cite{Israel92}.  The amplitude of the
fluid density profiles gradually decays away from the substrate to a
uniform bulk density profile.    Similarly, in the case of a
spherical particle, the spatial distribution of fluid monomers
around the particle consists of several concentric
layers~\cite{Schmidt04}.   In our simulations, the radial
distribution function for fluid monomers is shown in
Fig.\,\ref{fig:density_radial} for uniformly wetting
($\varepsilon_{\rm pf}=1.0\,\varepsilon$) and uniformly nonwetting
($\varepsilon_{\rm pf}=0.1\,\varepsilon$) particles.   In both
cases, the fluid density profiles decay to a uniform profile within
about $(3-4)\,\sigma$ from the particle surface.   As expected,
fluid layering is more pronounced for the uniformly wetting
particle.    In simulations with a Janus particle, a reduced value
of the surface energy at the nonwetting hemisphere resulted in a
lower amplitude of the local fluid density profile above the
nonwetting side, while the density oscillations remain large above
the wetting hemisphere where the interaction energy is fixed to
$\varepsilon$ (see the inset in Fig.\,\ref{fig:density_radial}).

\vskip 0.05in


The flow boundary conditions at the particle surface depend on the
atom density and the LJ interaction energy between particle atoms
and fluid monomers.    In order to estimate the Navier slip length,
we performed a set of separate simulations of fluid flow (without a
particle) confined between flat crystalline walls with the same
density as the Janus particle.   In this setup, the fluid phase with
density $\rho=0.75\,\sigma^{-3}$ is in contact with (111) planes of
the face-centered cubic (fcc) lattice.    Similar to the previous
studies~\cite{Priezjev07,PriezjevJCP}, the Poiseuille flow was
induced by a constant force $f_{x}=0.005\,\varepsilon/\sigma$
applied to each fluid monomer in the direction parallel to the
walls.    The slip lengths were computed from a parabolic fit of the
averaged velocity profiles for different values of the wall-fluid
interaction energy.    The results are summarized in
Table\,\ref{tabela}.      It is evident that at high interaction
energies, the slip length is about the particle radius, while at low
energies the slip length becomes larger than the particle diameter.
Note also that for flows over curved surfaces the slip length should
be modified by the local radius of
curvature~\cite{Panzer90,Priezjev06,Niavarani10,Koplik14}.

\begin{table}[t]
\caption{The slip length $L_s/\sigma$ computed in a Poiseuille flow
confined by flat crystalline walls with density $1.63\,\sigma^{-3}$
for the tabulated wall-fluid interaction energies.    The error bars
for the slip length are about $0.5\,\sigma$. }
 \vspace*{3mm}
 \begin{ruledtabular}
 \begin{tabular}{r r r r r r}
     \\ [-18pt]
     $\varepsilon_{\rm pf}/\varepsilon$ & $0.1$ & $0.3$ & $0.5$ & $0.7$ & $1.0~$
     \\ [5pt] \hline \\[-15pt]
     $L_s/\sigma$  &                      $9.0$ & $6.4$ & $4.6$ & $3.4$ & $2.0~$
     \\ [5pt]
 \end{tabular}
 \end{ruledtabular}
 \label{tabela}
\end{table}

\vskip 0.05in

\begin{figure}[t]
\includegraphics[width=12.cm,angle=0]{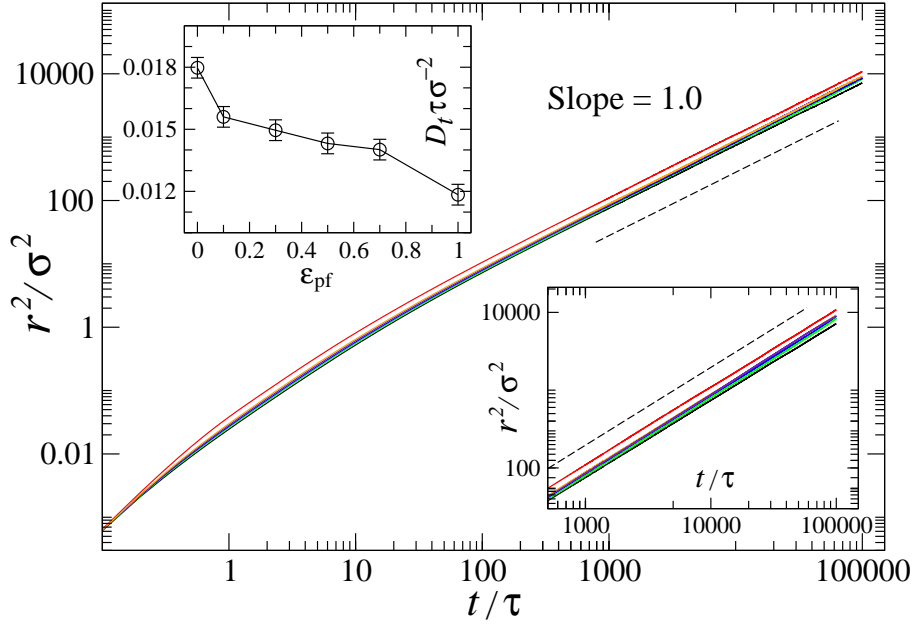}
\caption{(Color online)   The mean square displacement of particles
as a function of time for the surface energies at both hemispheres
from top to bottom ($0.1\,\varepsilon$, $0.1\,\varepsilon$),
($1.0\,\varepsilon$, $0.1\,\varepsilon$), ($1.0\,\varepsilon$,
$0.3\,\varepsilon$), ($1.0\,\varepsilon$, $0.5\,\varepsilon$),
($1.0\,\varepsilon$, $0.7\,\varepsilon$), and ($1.0\,\varepsilon$,
$1.0\,\varepsilon$).       The dashed line with unit slope is
plotted for reference.  The lower inset shows an enlarged view of
the same data at large times.    The upper inset shows the diffusion
coefficient as a function of $\varepsilon_{\rm pf}$, while the other
hemisphere is wetting, i.e., $\varepsilon_{\rm
pf}=1.0\,\varepsilon$.  The case $\varepsilon_{\rm pf}=0$
corresponds to a uniformly nonwetting particle ($0.1\,\varepsilon$,
$0.1\,\varepsilon$). } \label{fig:msd_time}
\end{figure}

In a quiescent solvent, the Janus particle undergoes translational
and rotational diffusion under random forces from the fluid
monomers.    We start the discussion of our results by considering
the position vector of the particle center of mass.  The mean square
displacement (MSD) for uniformly (non)wetting and Janus particles is
plotted in Fig.\,\ref{fig:msd_time} for different surface energies.
In all cases, the center of mass of the fluid phase was subtracted
from the particle position.    It can be clearly seen that after the
ballistic regime, the MSD increases linearly with time (as indicated
by the dashed line in Fig.\,\ref{fig:msd_time}).   As expected, the
slowest and fastest diffusion correspond to \textit{uniformly}
wetting and nonwetting particles, respectively. In between, the
displacement increases with decreasing surface energy at the
nonwetting hemisphere of Janus particles.   Overall, this result is
not surprising and can be expected for any spatial distribution of
binary wettability on the particle surface.  The translational
diffusion coefficient was computed from the relation $r^2=6D_t\,t$
at times $t\gtrsim10^3\,\tau$ (shown in the upper inset of
Fig.\,\ref{fig:msd_time}).

\vskip 0.05in

\begin{figure}[t]
\includegraphics[width=10.cm,angle=0]{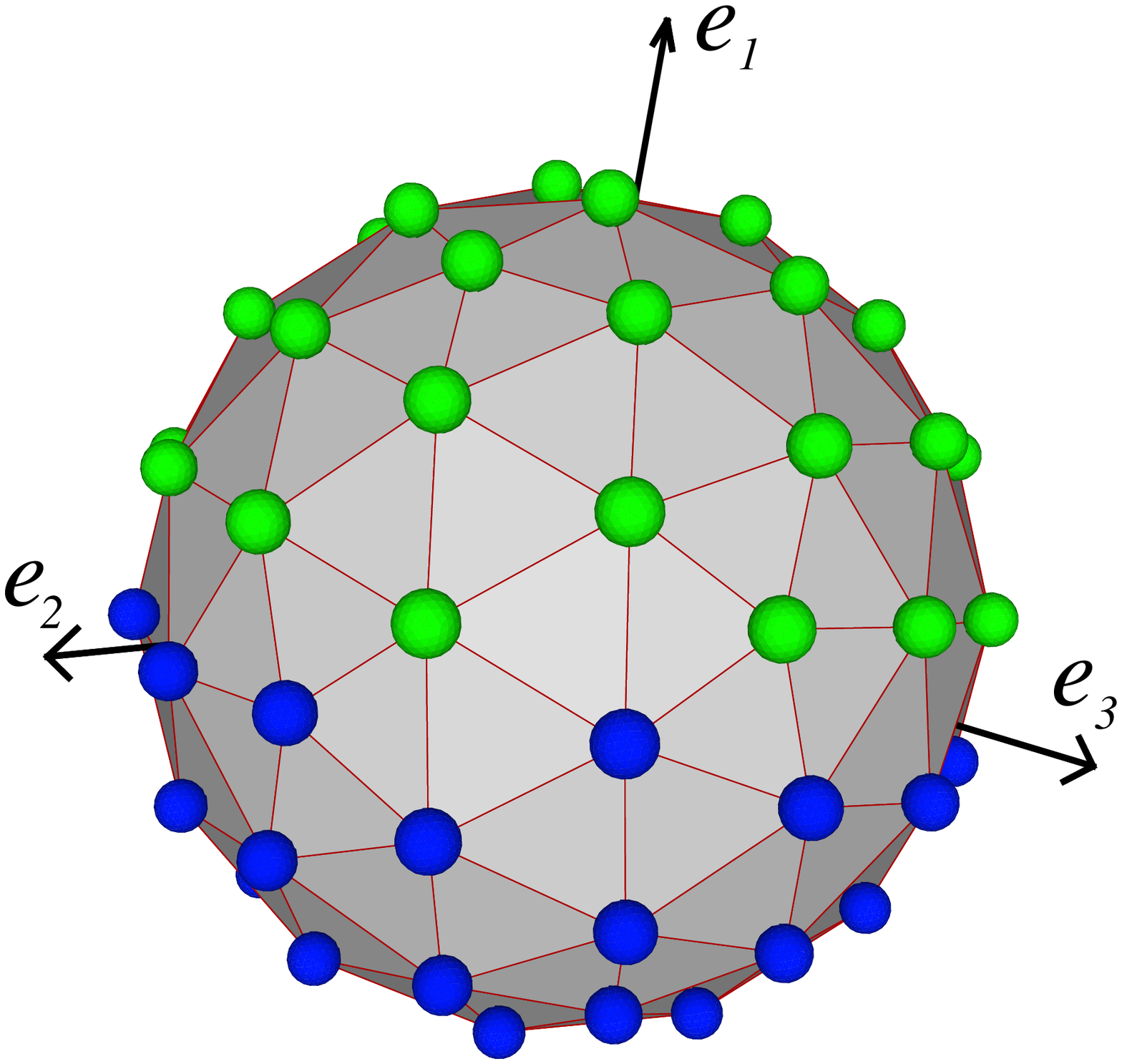}
\caption{(Color online) A snapshot of the Janus particle with
wetting (blue atoms) and nonwetting (green atoms) hemispheres. The
particle consists of $72$ atoms rigidly fixed on vertices of a
convex polyhedron (see text for details).    Note that particle
atoms are not drawn to scale.    The unit vectors $\textbf{e}_1$,
$\textbf{e}_2$ and $\textbf{e}_3$ form an orthonormal basis in a
reference frame of the particle. } \label{fig:snapshot_janus}
\end{figure}
\begin{figure}[t]
\includegraphics[width=12.cm,angle=0]{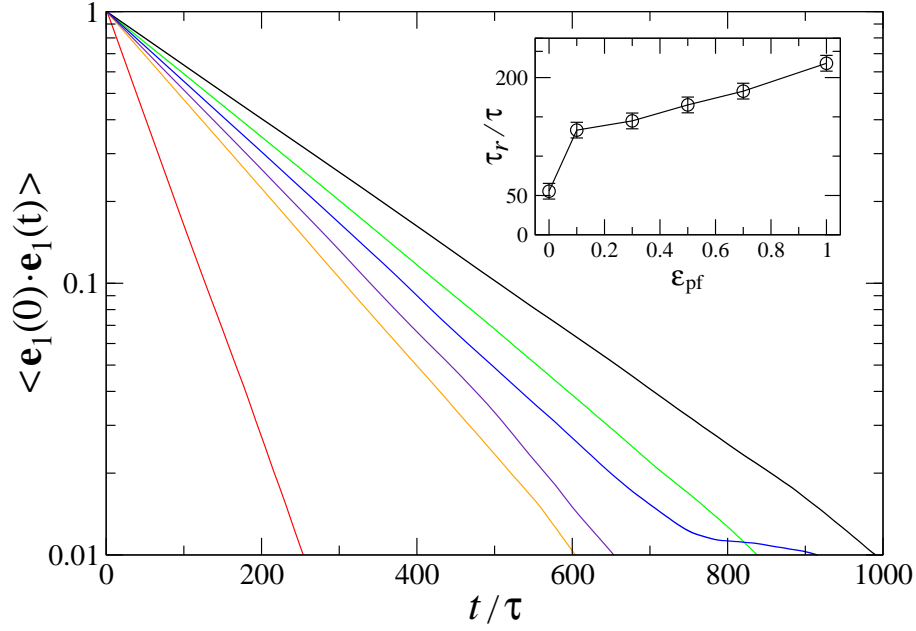}
\caption{(Color online)   The autocorrelation function $\langle
\textbf{e}_1(0)\cdot\textbf{e}_1(t)\rangle$ for the surface energies
at both hemispheres ($0.1\,\varepsilon$, $0.1\,\varepsilon$),
($1.0\,\varepsilon$, $0.1\,\varepsilon$), ($1.0\,\varepsilon$,
$0.3\,\varepsilon$), ($1.0\,\varepsilon$, $0.5\,\varepsilon$),
($1.0\,\varepsilon$, $0.7\,\varepsilon$), and ($1.0\,\varepsilon$,
$1.0\,\varepsilon$) from left to right.   The inset shows the
rotational relaxation time $\tau_r$ (in units of $\tau$) as a
function of the surface energy at the nonwetting hemisphere. }
\label{fig:e1e1_time}
\end{figure}

The rotational dynamics of nanoparticles can be probed by analyzing
the decay of the autocorrelation function of the unit vectors
$\textbf{e}_i$ that are fixed relative to the particle.   In what
follows, we choose the vector $\textbf{e}_1$ along the main axis of
symmetry of the particle (see Fig.\,\ref{fig:snapshot_janus}). The
autocorrelation function $\langle \textbf{e}_1(0) \cdot
\textbf{e}_1(t) \rangle$ is plotted in Fig.\,\ref{fig:e1e1_time} for
the Janus and uniformly wetting and nonwetting particles.  One can
observe a similar trend as with the translational diffusion; namely,
with increasing surface energy at the nonwetting hemisphere, the
rotational relaxation slows down, and the two limiting cases of
uniformly (non)wetting particles correspond to (fastest) slowest
decay.  A typical rotational relaxation time, $\tau_r$, was
estimated from an fit to a simple exponential decay, i.e.,
\begin{equation}
\langle \textbf{e}_1(0) \cdot \textbf{e}_1(t) \rangle =
\textrm{exp}(-t/\tau_r). \label{Eq:e1e1}
\end{equation}
The results are reported in the inset of Fig.\,\ref{fig:e1e1_time}.
It can be observed that $\tau_r$ gradually increases with increasing
surface energy $\varepsilon_{\rm pf}$, which implies that the
rotational diffusion coefficient is larger for Janus particles with
lower surface energy at the nonwetting hemisphere.


\vskip 0.05in

\begin{figure}[t]
\includegraphics[width=12.cm,angle=0]{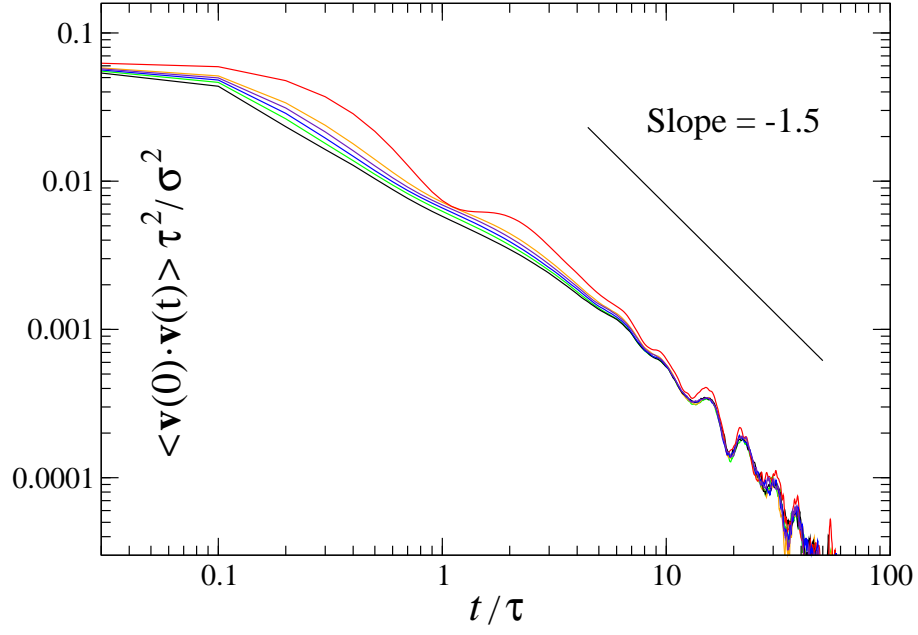}
\caption{(Color online) The velocity autocorrelation function
$\langle \textbf{v}(0)\cdot\textbf{v}(t) \rangle$ for the
interaction energies at both hemispheres ($0.1\,\varepsilon$,
$0.1\,\varepsilon$), ($1.0\,\varepsilon$, $0.1\,\varepsilon$),
($1.0\,\varepsilon$, $0.3\,\varepsilon$), ($1.0\,\varepsilon$,
$0.5\,\varepsilon$), ($1.0\,\varepsilon$, $0.7\,\varepsilon$), and
($1.0\,\varepsilon$, $1.0\,\varepsilon$) from top to bottom. The
straight line with a slope $-1.5$ is shown as a reference. }
\label{fig:vacf_time}
\end{figure}
\begin{figure}[t]
\includegraphics[width=12.cm,angle=0]{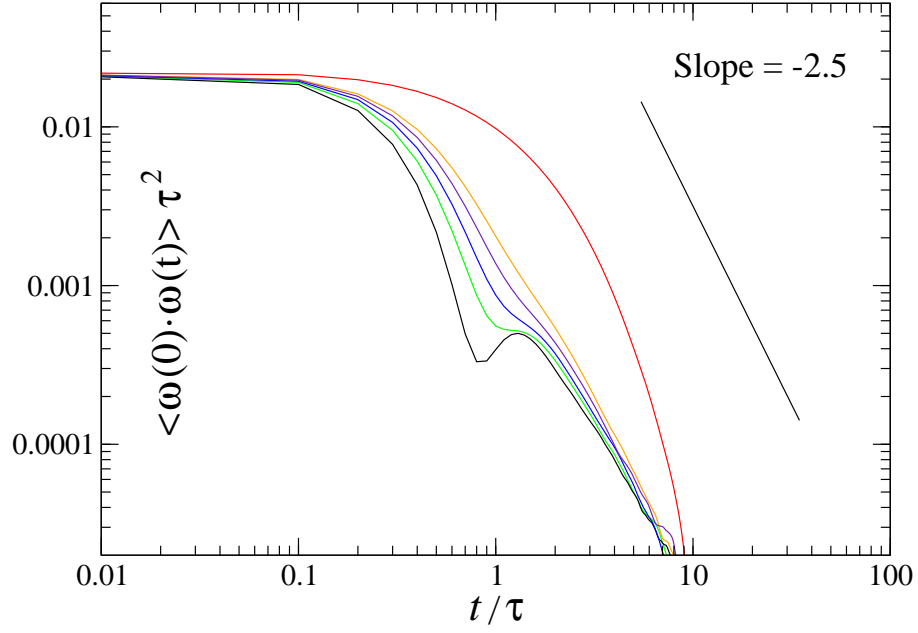}
\caption{(Color online) The angular velocity autocorrelation
function $\langle\boldsymbol\omega(0)\cdot\boldsymbol\omega(t)
\rangle$ for the LJ interaction energies at both hemispheres
($0.1\,\varepsilon$, $0.1\,\varepsilon$), ($1.0\,\varepsilon$,
$0.1\,\varepsilon$), ($1.0\,\varepsilon$, $0.3\,\varepsilon$),
($1.0\,\varepsilon$, $0.5\,\varepsilon$), ($1.0\,\varepsilon$,
$0.7\,\varepsilon$), and ($1.0\,\varepsilon$, $1.0\,\varepsilon$)
from top to bottom. The black line with a slope $-2.5$ is plotted
for reference. } \label{fig:avacf_time}
\end{figure}

In order to emphasize the difference between uniform and anisotropic
particles, we next consider the translational and angular velocity
autocorrelation functions (VACF) defined as $\langle
\textbf{v}(0)\cdot\textbf{v}(t)\rangle$ and $\langle
\boldsymbol\omega(0)\cdot\boldsymbol\omega(t)\rangle$, respectively.
First, we plot the translational VACF in Fig.\,\ref{fig:vacf_time}
for uniformly (non)wetting and Janus particles.    Note that in all
cases, the VACF is determined by the thermal velocity
$\textbf{v}^2=3\,k_BT/M$ at short times, while the algebraic decay
with the slope $-1.5$ is recovered at long times, which is
consistent with the previous MD
results~\cite{Alder70,Levesque07,Chakraborty11}.      In the
intermediate regime, the VACF is smaller for Janus particles with
higher interaction energy at the nonwetting hemisphere.   Similarly,
the angular VACF for Janus particles gradually varies between the
two limiting cases of uniformly wetting and nonwetting particles
(shown in Fig.\,\ref{fig:avacf_time}).     Although the statistics
at long times is limited, it appears that the angular VACFs exhibit
a power-law decay with the slope $-2.5$, in agreement with earlier
predictions~\cite{Berne71,Davis75}.

\vskip 0.05in


The autocorrelation functions discussed above do not necessarily
reflect the broken symmetry of Janus particles; in principle, one
can expect an enhanced diffusion with decreasing surface wettability
regardless of its spatial distribution.   In what follows, we
explore a correlation between translational and rotational motion of
Janus particles.    It is intuitively clear that if a constant force
is applied to a Janus particle in a viscous fluid then the
nonwetting face will be oriented along the direction of motion in
order to reduce drag.    In analogy, when a Janus particle moves
under random forces from the surrounding fluid, its nonwetting
hemisphere tends to rotate in the direction of the displacement
vector.    This effect can be quantified by computing a scalar
product of the displacement vector of the center of mass during the
time interval $t$ and a vector difference between unit vectors
$\textbf{e}_1(t)$ and $\textbf{e}_1(0)$, and then plotting it as a
function of time (not shown).    However, we found that it is
difficult to obtain good statistics for this quantity and, more
importantly, to provide a clear physical interpretation of the
effect.

\vskip 0.05in

\begin{figure}[t]
\includegraphics[width=12.cm,angle=0]{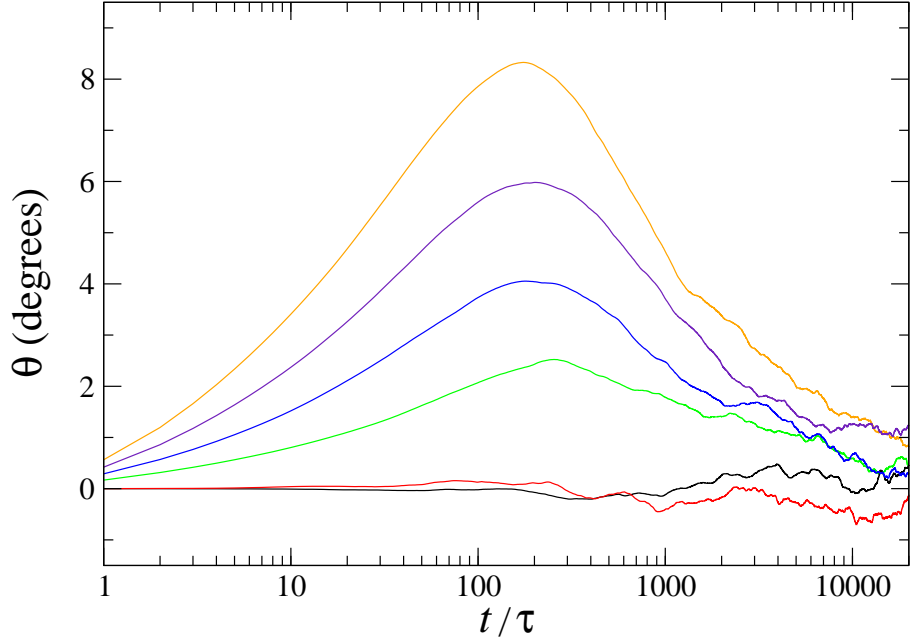}
\caption{(Color online) The rotation angle along the displacement
vector of the center of mass for the surface energies at both
hemispheres ($1.0\,\varepsilon$, $0.1\,\varepsilon$),
($1.0\,\varepsilon$, $0.3\,\varepsilon$), ($1.0\,\varepsilon$,
$0.5\,\varepsilon$), ($1.0\,\varepsilon$, $0.7\,\varepsilon$) from
top to bottom.   The data for uniformly wetting ($1.0\,\varepsilon$,
$1.0\,\varepsilon$) and nonwetting ($0.1\,\varepsilon$,
$0.1\,\varepsilon$) particles are indicated by the black and red
curves, respectively.  } \label{fig:angle}
\end{figure}

Alternatively, a more direct way to probe the correlation between
translation and orientation of Janus particles is to compute a
rotation angle along the direction of displacement.    More
specifically, we first choose a time interval $t$, compute the
displacement vector of the center of mass $\triangle\textbf{r}$,
determine angles between ($\textbf{e}_1(0)$, $\triangle\textbf{r}$)
and ($\textbf{e}_1(t)$, $\triangle\textbf{r}$), and then plot the
difference between these angles as a function of time.   In other
words, we measure how the unit vector $\textbf{e}_1$ rotates along
but not around the displacement vector.    The averaged rotation
angle is plotted in Fig.\,\ref{fig:angle} as a function of time for
the Janus and uniformly wetting and nonwetting particles.     It can
be clearly observed that, within statistical uncertainty, particles
with uniform distribution of wettability do not, on average, rotate
along the displacement vector.     In contrast, Janus particles
undergo a finite rotation, which increases with decreasing surface
energy at the nonwetting hemisphere.   In each case, the rotation
angle is maximum at times of about the rotational relaxation time
(see inset in Fig.\,\ref{fig:e1e1_time}).   Note also that the
maximum rotation angle in Fig.\,\ref{fig:angle} occurs at larger
times for Janus particles with higher surface energy, which
correlates with a slower relaxation of the autocorrelation function
$\langle \textbf{e}_1(0)\cdot\textbf{e}_1(t)\rangle$ for Janus
particles with higher $\varepsilon_{\rm pf}$ reported in
Fig.\,\ref{fig:e1e1_time}.   As expected, at longer times the
correlation gradually decays, although it remains nonzero even at
times of about $10^4\,\tau$.

\vskip 0.05in

\begin{figure}[t]
\includegraphics[width=12.cm,angle=0]{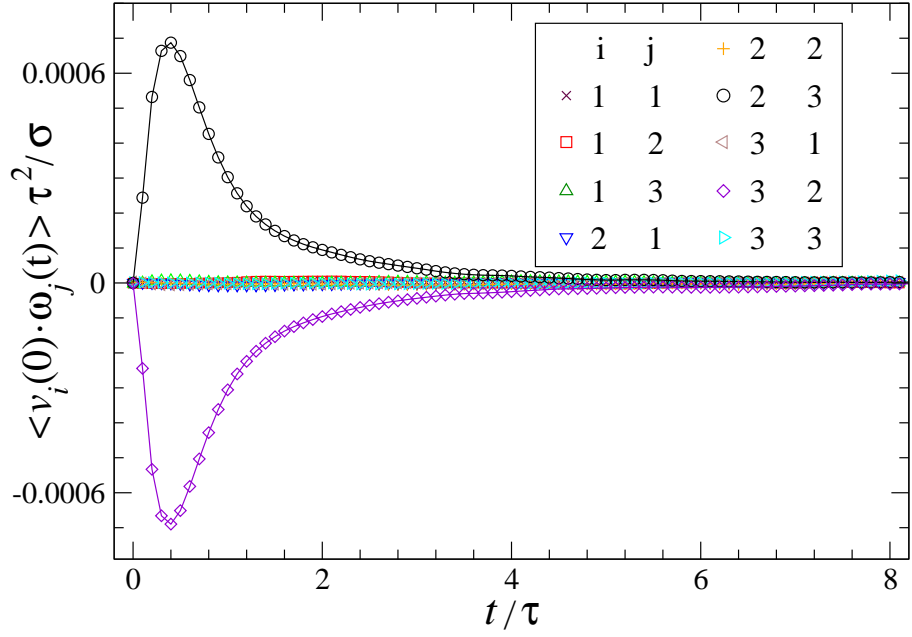}
\caption{(Color online) The translational times angular velocity
correlation functions $\langle v_i(0)\cdot\omega_j(t)\rangle$ for
the indicated pairs of indices.  The velocity components are
computed along the unit vectors $\textbf{e}_i$ (see
Fig.\,\ref{fig:snapshot_janus}). The data are reported for a Janus
particle with the LJ interaction energies at both hemispheres
($1.0\,\varepsilon$, $0.1\,\varepsilon$).   } \label{fig:cross_corr}
\end{figure}

To further quantify the effect of rotation along the displacement,
we also measured a correlation between the translational and angular
velocities.     The corresponding correlation functions were defined
as a product $\langle v_i(0) \cdot \omega_j(t)\rangle$, where the
velocity components with indices $i, j=1, 2, 3$ were computed along
the unit vectors $\textbf{e}_i$ (see
Fig.\,\ref{fig:snapshot_janus}). As a representative case, we
consider a Janus particle with the largest difference in
wettability, ($1.0\,\varepsilon$, $0.1\,\varepsilon$), when the
correlation effect is most pronounced.      The averaged correlation
functions are presented in Fig.\,\ref{fig:cross_corr} for selected
pairs of indices.    It can be seen that the only non-zero
correlation functions are $\langle v_2(0) \cdot \omega_3(t)\rangle$
and $\langle v_3(0) \cdot \omega_2(t)\rangle$.  Physically, it means
that if a Janus particle has a translational velocity component in a
direction parallel to the symmetry plane then it will tend to rotate
its nonwetting hemisphere towards the displacement. Finally, we note
that for uniformly wetting/nonwetting particles there is no
preferred rotation with respect to translation, and, therefore, the
cross correlation functions are zero at all times (not shown).

\section{Conclusions}

In summary, we investigated the translational and rotational
diffusion of individual Janus nanoparticles in a dense fluid using
molecular dynamics simulations.    In our model, the Janus particle
consists of atoms arranged on a sphere, which is divided in two
hemispheres of different wettability.      The interaction energy
between particle atoms and fluid molecules is fixed at the wetting
hemisphere, while the wettability of the other side is a variable
parameter.      The diffusion process of such particles is compared
against two limiting cases of uniformly wetting and nonwetting
particles.

\vskip 0.05in

We found that the diffusion coefficient of Janus particles is
bounded between the two limiting cases and it increases with
decreasing surface energy at the nonwetting hemisphere.   The
analysis of the translational and angular velocity autocorrelation
functions showed that the exponents of the long-time power-law decay
are the same for Janus and uniform particles.   Perhaps most
interestingly, we observed a correlation between translation and
rotation of Janus particles; in other words, when a particle
spontaneously moves in a particular direction it tends to reduce
drag by rotating its nonwetting side towards the displacement
vector.  This effect is absent for homogeneous particles.

\vskip 0.05in

In the future, it will be instructive to extend these results to a
system of many particles and study in detail the dynamics of
self-assembly of Janus particles into complex structures in the bulk
or at liquid-liquid interfaces.

\section*{Acknowledgments}

Financial support from the National Science Foundation
(CBET-1033662) is gratefully acknowledged.  Computational work in
support of this research was performed at Michigan State
University's High Performance Computing Facility and the Ohio
Supercomputer Center.  The molecular dynamics simulations were
conducted using the LAMMPS numerical code~\cite{Lammps}.



\bibliographystyle{prsty}

\end{document}